----------
X-Sun-Data-Type: default
X-Sun-Data-Description: default
X-Sun-Data-Name: geometry.tex
X-Sun-Charset: us-ascii
X-Sun-Content-Lines: 323

\documentstyle[12pt]{article}
\newcommand{\p}[1]{(\ref{#1})}
\begin{document}
\thispagestyle{empty}
\renewcommand{\thefootnote}{*}

\hbox to \hsize{\hfill KCL-TH-94-9}
\vskip1.5truecm
\centerline
{\bf GEOMETRY OF FERMIONIC CONSTRAINTS}
\centerline
{\bf IN SUPERSTRING THEORIES\footnote{Talk given at the Conference on Geometry
of Constrained Dynamical Systems,
Cambridge, June 15--18, 1994.}}

\vskip 1.5truecm
\centerline
{\bf
Dmitrij P. Sorokin
\renewcommand{\thefootnote}{\dagger}
\footnote{Permanent address: Kharkov Institute of Physics and Technology,
	Kharkov, 310108, the Ukraine,}\renewcommand{\thefootnote}{}
\footnote{e-mail address: dsorokin@kfti.kharkov.ua}}

\bigskip
\centerline{\sl Math. Department, King's College, Strand, London, WC2R 2LS, UK}

\vskip 1.5truecm
\renewcommand{\thefootnote}{\arabic{footnote}}
\setcounter{footnote}0

During recent years there has been an activity in the development of a,
so called, twistor-like, doubly supersymmetric approach for describing
superparticles and superstrings \cite{stv}--\cite{dg}. The aim of the  approach
is to provide with clear geometrical meaning an obscure local fermionic
symmetry      ($\kappa$--symmetry) of
superparticles and superstrings \cite{ks,gsw}, which plays an essential role in
quantum consistency of the theory. At the same time this local fermionic
symmetry causes problems with performing the covariant Hamiltonian analysis and
quantization of the theories. This is due to the fact that
the first--class constraints corresponding to the $\kappa$--symmetry form an
infinit reducible set, and in a conventional formulation of superparticles and
superstrings (see \cite{gsw} and references therein) it turned out impossible
to single out an irreducible set of the fermionic first--class constraints in a
Lorentz covariant way.
So the idea was to replace the $\kappa$--symmetry by a local extended
supersymmetry on the worldsheet by constructing superparticle and superstring
models which would be manifestly supersymmetric in a target superspace and on
the worldsheet with the number of local  supersymmetries being equal to
the number of independet $\kappa$--symmetry transformations, that is
$n=D-2$ in a space--time with the dimension D=3, 4, 6 and 10. Note that it is
just in these space--time dimensions the classical theory of Green--Schwarz
superstrings may be formulated \cite{gsw}, and twistor relations \cite{lli}
take place.

The doubly supersymmetric formulation provides the ground for natural
incorporating twistors into the structure of supersymmetric theories. Twistor
components (which are roughly speaking commuting spinors) arise as
superpartners of Grassmann spinor coordinates of the target superspace and
allow one to solve
such problems of superparticles and superstrings as the geometrical nature of
the fermionic $\kappa$--symmetry, the Lorentz--covariant separation of the
first and second class constraints, and finding the way of establishing, at the
classical level, the relationship between Green--Schwarz and the
Neveu--Schwarz--Ramond formulation of the strings, and the hope is that the
twistor--like approach may contain the advantages of the both these
formulations.

In this talk I would like to present basic ideas of the twistor--like approach
developed so far.

Consider a conventional action describing the dynamics of a massless $N=1$
superparticle in
space-time superspace parametrized by bosonic vector coordinates $x^m$ and
fermionic
spinor coordinates $\theta^\alpha$ (m=0,1,...,D-1; $\alpha$=1,...,2D-4;
D=3,4,6,10). In the first order form the action looks as follows:
\begin{equation}\label{1}
S=\int d\tau [p_m({d\over{d\tau}}x^m-i{d\over{d\tau}}\bar\theta\gamma^m\theta)
-{1\over 2}e(\tau)p_mp^m],
\end{equation}
where $e(\tau)$ is an auxiliary (one-dimensional gravity) field which ensures
the
momentum $p_m$ of the superparticle to be light--like on the mass shell:
\begin{equation}\label{2}
p_mp^m=0.
\end{equation}

The action is invariant under the reparametrization of the time parameter
$\tau\to f(\tau)$, N=1 target space supersymmetry:
\begin{equation}\label{3}
\delta\theta=\epsilon,\qquad \delta x^m=-i\delta\bar\theta\gamma^m\theta
\end{equation}
and local fermionic $\kappa$--symmetry:
\begin{equation}\label{3.}
\delta p_m=0,\qquad
\delta\theta_\alpha=i(p_m\gamma)_\alpha^{m\beta}\kappa_\beta(\tau),\qquad
\delta x^m=i\delta\bar\theta\gamma^m\theta,\qquad \delta
e=4\bar\kappa\dot\theta,
\end{equation}
where $\kappa_\alpha(\tau)$ is a 2(D-2)--component parameter of the
transformations, while, due to the lightlikeness of $p_m$ \p{2}, only n=D-2
$\kappa$--symmetry parameters are independent.

Let us replace the $\kappa$--symmetry with a local n=D-2 extended supersymmetry
on the worldline of the superparticle. To this end we have to construct a
version of the theory being manifestly invariant under the worldline
supersymmetry. It means that we shall consider the trajectory of the
superparticle
to be a  worldline superspace parametrized by $\tau$ and Grassmann coordinates
$\eta^a$ (a=1,...,D-2), and the target space coordinates
$X^m(\tau,\eta)=x^m(\tau)+\eta^a\chi^m_a(\tau)+...;~~
\Theta_\alpha(\tau,\eta)=\theta_\alpha(\tau)+\eta^a\lambda_a(\tau)+...$ of the
superparticle to be superfields in the worldline superspace. Thus, we see that
$\theta(\tau)$ acquires commuting spinors $\lambda_a$ as its superpartners.

The generalization of eq.\p{1} to the case D=3, N=1, n=1 is straightforward.
Instead of the time derivative we use supercovariant derivative
$D={\partial\over{\partial\eta}}+i\eta{\partial\over{\partial\tau}}$ and write
down a generalized action in the following form \cite{pas}:
\begin{equation}\label{4}
S=\int d\tau d\eta[P_m(\tau,\eta)(DX^m-iD\bar\Theta\gamma^m\Theta)-{1\over 2}
E(\tau,\eta)P_mP^m].
\end{equation}

Action \p{4} is invariant under $\tau$--reparametrization; n=1 local
supersymmetry transformations, which, in particular, for $\Theta(\tau,\eta)$
components look as follows:
\begin{equation}\label{5}
\delta\theta_\alpha=\alpha(\tau)\lambda_\alpha;
\end{equation}
and fermionic transformations \p{3.}, where all variables are replaced by
corresponding superfields, denoted by capital letters, and ${d\over{d\tau}}$ is
replaced by the supercovariant derivative $D$. At the first glance it seems
that we have not
got rid of the $\kappa$--symmetry since it  appeared again at the  superfield
level. But, in addition, action \p{4} is invariant also under the following
bosonic superfield transformations
\begin{equation}\label{6}
\delta X^m=\Lambda(\tau,\eta)P^m, \qquad \delta E=D\Lambda.
\end{equation}
And it turns out that the transformations \p{6} and the superfield
generalization of \p{3.} allows one to gauge fix $E(\tau,\eta)$ to be zero
{\sl globally} in the worldline superspace \cite{pas}. Thus, the last term
drops out of
the eq.\p{4}, and we get the action originally obtained in \cite{stv}:
\begin{equation}\label{7}
S=\int d\tau d\eta P_m(\tau,\eta)(DX^m-iD\bar\Theta\gamma^m\Theta),
\end{equation}
which possesses only the doubly supersymmetry.

Integrating \p{7} over $\eta$ and eliminating auxiliary fields one arrives at
a component action
\begin{equation}\label{8}
S=\int d\tau p_m({d\over{d\tau}}x^m-i{d\over{d\tau}}\bar\theta\gamma^m\theta
-\bar\lambda\gamma^m\lambda).
\end{equation}
 As the solution to the equation of motion of $\lambda$ we get the twistor
representation of the light--like vector in D=3,4,6 and 10 space--time
dimensions:
\begin{equation}\label{9}
p_m\gamma^{m\beta}_\alpha\lambda_\beta=0 \hskip24pt\rightarrow \hskip24pt
p^m\sim\bar\lambda\gamma^m\lambda \hskip24pt \rightarrow \hskip 24pt p_mp^m=0.
\end{equation}
Substituting the twistor representation of $p_m$ into \p{3.} one may convince
oneself that the $\kappa$ transformations coincide (for D=3) with the local
supersymmetry
transformations \p{5} with $\alpha(\tau)=\bar\lambda\kappa$, and it can be
shown that the model is equivalent to the conventional N=1 superparticle
\cite{stv}.

There are differenet ways of generalizing action \p{7} to the case of
D=4,6,10 and n=D-2 \cite{stv,ds,n8p,pas}. The most straightforward (and the
only known for D=10)
 generalization is achieved by extending the number of Grassmann coordinates
$\eta^a$ and writing down the action in the form \cite{n8p}:
\begin{equation}\label{10}
S=\int d\tau d^{D-2}\eta
P_{ma}(\tau,\eta)(D_{a}X^m-iD_{a}\bar\Theta\gamma^m\Theta).
\end{equation}
The nontrivial thing is to show that in spite of a rather rich contents of the
superfields in this action there are enough local symmetries and equations of
motion to kill all auxiliary fields so that the model is classiclally
equivalent
to the masless Brink--Schwarz superparticle \cite{n8p}.

The next step is to generalize this doubly supersymmetric action to $N=1$
superstrings. For this we suppose that a hypersurface swept by the string is a
worldsheet superspace with heterotic geometry subject to constraints on torsion
\cite{h,t}, and the points of this surface are parametrized by ($\tau, \sigma,
\eta^{-}_a$). Again, the straightforward generalization of \p{10} is
\begin{equation}\label{11}
S=\int d\tau d^{D-2}\eta
P_{ma}(\tau,\eta)(D_{-a}X^m-iD_{-a}\bar\Theta\gamma^m\Theta),
\end{equation}
where in a Wess--Zumino gauge $D_{-a}={\partial\over{\partial\eta^{_a}}}+
i\eta^-_a e^\mu_{--}(\xi){\partial\over{\partial\xi^{\mu}}}$ with
$e^\mu_{--}(\xi)$ being one of the two worldsheet zweinbeins $(\xi^{\mu}=(\tau,
\eta)).$

It can be shown \cite{bstv} that this action describes so called N=1 null
superstring, that is a string with zero tension and a degenerate worldsheet
metric \cite{nul}. This null superstring is infact a continuous set of massless
superparticles moving in such a way that their momenta are orthogonal to the
string so that the null string does not fall into pieces. The action \p{11},
as well as the ones for superparticles, was called a geometro--dynamical term
\cite{gs2}, since it determines the dynamics of a superstring by specifying the
imbedding of superworldsheet into target superspace. This imbedding is
characterized by vanishing components of the one form
$\Pi^m=dX^m-id\bar\Theta\gamma^m\Theta$ along the Grassmann--odd directions of
the worldsheet superspace:
\begin{equation}\label{12}
{\delta S\over{\delta P_{ma}}}=D_{a}X^m-iD_{a}\bar\Theta\gamma^m\Theta=0.
\end{equation}
Thus, if the dynamics of a superstring is described solely by the
geometro--dynamical term \p{11} it is profitable for the string to propagate as
the null superstring.

To get a fully fledged superstring we have to further specify the imbedding in
a way which leads to string tension generation \cite{ten,dg}. This is achieved
by
determining the pullback of a Wess--Zumino  two form
\begin{equation}\label{13}
B=i\Pi^m\wedge d\bar\Theta\gamma_m\Theta.
\end{equation}
In the conventional Green--Schwarz approach the pullback of this form on a
two--dimensional worldsheet is a closed form as any two form on a
two--dimensional manifold.

We would like this property to be valid in our case
as well. But now getting the closure of the pullback of the form $B$ is not a
trivial problem anymore, since the worldsheet is a supermanifold.

First of all the twistor condition \p{12} on the one form $\Pi^m$ must be
satisfied.

Secondly, even then we have to modify the form $B$ in the following way
\cite{dg}:
\begin{equation}\label{14}
\hat B=B-{1\over{D-2}}(E^{--}\wedge
E^{++})D_{-a}(\bar\Theta\gamma_mD_{-a}\Theta)
E^M_{++}(\partial_MX^m-i\partial_M\bar\Theta\gamma^m\Theta),
\end{equation}
where $E^{\pm\pm}_M$ are components of the worldsheet supervielbeins
($M,N=\mu,a $).

Thus, porvided the twistor condition \p{12} is valid, one may check that $\hat
B$ is a closed form in the worldsheet superspace. It means that external
differential of $\hat B$ is zero, and locally $\hat B$ is an exact form:
\begin{equation}\label{15}
\hat B=dA,
\end{equation}
where $A_M(\xi,\eta)$ is an ``electromagnetic'' superfield on the worldsheet
\cite{ten}.

It is desirable to get this condition as one determined by the dynamics of the
string. To this end let us add to the string action \p{11} a term from which
this condition can be obtained:
\begin{equation}\label{16}
S_{wz}=\int d^2\xi d^{D-2}\eta P^{MN}(\hat B_{MN}-\partial_M A_N),
\end{equation}
where $P^{MN}=(-1)^{MN+1}P^{NM}$ is a Grassmann antisymmetric Lagrange
multiplier.
Action \p{16} together with \p{11} is invariant under the following
transformations of $P^{MN}$:
\begin{equation}\label{17}
\delta P^{MN}=\partial_L\Lambda^{LMN}(\xi,\eta),
\end{equation}
where $\Lambda^{LMN}$ is a Grassmann antisymmetric superfield parameter.
The equation of motion of $A_N$ gives
$$
\partial_MP^{MN}=0.
$$
The solution to this equation, with taking into account gauge fixing for the
transformations of $P^{MN}$ \p{17}, reads as follows.
$$
p^{\mu\nu}=\varepsilon^{\mu\nu}T\eta^{D-2},
$$
while other components of the superfield $P^{MN}$ are zero. $T$ is a constant
which is identified with string tension.

When $T=0$ we again get the null superstring, since then \p{16} vanishes. When
$T\not =0$, one may eliminate in \p{16}
all auxiliary degrees of freedom and get the conventional \hbox{$N=1$}
Green--Schwarz
superstring action, which is a target--space supersymmetric part of a
heterotic string action \cite{gsw}.

To complete the  doubly supersymmetric formulation of the heterotic string one
has to construct a superfield action for describing chiral fermions
which must be taken into account for the quantum consistency of the theory.
Two possible ways of how one may try to do this has been proposed
\cite{st,howe}, but
the problem was not completely solved, since either there is a danger
that undesirable auxiliary degrees of freedom may become propagative \cite{st},
or
an internal gauge group associated with the chiral fermions is too small
\cite{howe}. Recently a modified version of \cite{st} was proposed in
\cite{is}, where,
as the authors argue, both these problems are solved.

The program of doubly supersymmetric twistorization has been fulfilled for
superparticles and superstrings in D=2,3,4,6 and 10 space--time dimensions with
N=1, and n=D-2, but the generalization of these results to an N=2
Green--Schwarz superstring encountered problems with not allowing auxiliray
fields of the model to propagate. Various versions of the twistor--like N=2
Green-Schwarz superstrings have been studied in \cite{vz,cp,pt}, and
twistor--like supermembrane models were constructed in \cite{ptm}. A model for
describing doubly supersymmetric
heterotic string with the both Virasoro constraints solved in the twistor form
\cite{comp}, and the existing versions of the chiral fermion action indicate
that one might hope to overcome the problem of propagating Lagrange
multipliers. Work in this direction is in progress.

\bigskip
{\bf Acknowledgments}. I would like to thank Nathan Berkovits, Paul Howe and
the members of the theoretical group for
their warm hospitality at Maths. Department of King's College, London, and
the Royal Society for awarding me a Kapitza Fellowship for visiting London and
Cambridge. This work was partially supported by the International Science
Foundation under the grant No RY 9000.

\end{document}